\documentclass[times, 10pt,twocolumn]{article} 
\usepackage{latex8}
\usepackage{times}
\usepackage{graphicx}
\usepackage{xcolor}
\usepackage{amsmath,amssymb,amsfonts}
\DeclareMathOperator{\E}{\mathbb{E}}
\usepackage{svg}
\usepackage{caption}
\usepackage{multirow}
\usepackage{url}  
\usepackage{subcaption}

\begin{document}

\title{Feature Compression for Rate Constrained Object Detection on the Edge}

\author{Zhongzheng Yuan
\qquad
Samyak Rawlekar \qquad Siddharth Garg \qquad Elza Erkip \qquad Yao Wang \\ \\
New York University Tandon School of Engineering \\ Electrical and Computer Engineering Department \\ 6 MetroTech Center, Brooklyn, New York 11201, USA \\ 
\{zy740, skr2369, siddharth.garg, elza, yaowang\}@nyu.edu \\
}

\maketitle
\thispagestyle{empty}

\begin{abstract}
    Recent advances in computer vision has led to a growth of interest in deploying visual analytics model on mobile devices.
    However, most mobile devices have limited computing power, which prohibits them from running large scale visual analytics neural networks. An emerging approach to solve this problem is to offload the computation of these neural networks to computing resources at an edge server.     Efficient computation offloading requires optimizing the trade-off between multiple objectives including compressed data rate, analytics performance, and computation speed. In this work, we consider a ``split computation" system to offload a part of the computation of the YOLO object detection model. We propose a  learnable feature compression approach to compress the intermediate YOLO features with light-weight computation. We train the  feature compression and decompression module together with the YOLO model to optimize the object detection accuracy under a rate constraint. Compared to baseline methods that apply either standard image compression or learned image compression at the mobile and perform image decompression and YOLO at the edge, the proposed system  achieves higher detection accuracy at the low to medium rate range. Furthermore, the proposed system requires substantially lower computation time on the mobile device with CPU only.  
\end{abstract}


\Section{Introduction}

Offloading visual analytics computations (such as object detection) from images captured by mobile devices to an edge server can reduce the computation time and the power consumption of the mobile device. Reduced computation time  is critical for real-time applications such as navigation and robot control, while reduced power consumption can extend the battery life of the mobile device. Research in this direction has proposed two broad approaches. In one, a mobile device directly compresses images which are then decompressed by the edge server for visual analytics \cite{choi2020task, chamain2021end, le2021image}. The second approach is to perform a part of the visual analytics task on the mobile device and compress intermediate features; 
the server then decompresses these features and completes the remainder of the analytics task~\cite{chen2019toward,duan2020video,eshratifar2019bottlenet,singh2020end,matsubara2022supervised}. This approach is commonly known as the ``split computation."

In either approach, compression can be performed using conventional, non-learnable compression methods. Such approaches may have practical advantages because they can leverage existing hardware and software for compression, but the impact of compression on the analytics task performance cannot be controlled directly. Using a learnable compression module allows one to directly optimize the rate-analytics trad-off.  Furthermore, the ``split-computation'' framework with learnable compression has the potential to achieve a better rate-analytics trade-off,  because it only needs to generate and compress the features that are useful for the analytics task. The split computation approach can also reduce the computation at the server, which could be important for applications where the server has resource constraints.

In this paper, we propose a ``split computation'' system with learnable feature compressor and decompressor.  To compress the multi-channel features at the split point of the YOLO model, we perform channel reduction to reduce the number of channels and furthermore decorrelate the resulting channels. We use the hyperprior idea of \cite{balle2018variational} to encode the reduced features by introducing additional hyperprior encoder and decoder. Our system is developed and evaluated for object detection using the YOLOv5 model architecture \cite{YOLOv5}. However, the general methodology is applicable for other visual analytics tasks and other object detection model architectures. 

We evaluate the performance of the proposed system for a common object detection task: detecting 80 object classes in the COCO dataset (testing set).
Compared to applying a state-of-the-art image compression method (BPG) \cite{BPG} or the learned image compression model optimized for image reconstruction \cite{balle2018variational}, followed by the pretrained YOLOv5 model on the decompressed images, the proposed method was able to achieve higher detection accuracy over a large rate range.  
Compared to  jointly refining the learned image compression  \cite{balle2018variational} and the YOLOv5 detection models, the proposed method was able to achieve better performance at the low to medium rate regime.  We further show that when only a limited set of objects (e.g. people, vehicle, traffic lights, etc.) are relevant for a specific application (e.g. traffic monitoring), the proposed approach provides substantial gain over the entire rate range, because the relevant features can be captured by a very small number of channels. In addition to the rate-detection trade-off, the computation complexity at the mobile device is a critical factor for mobile applications. 
We demonstrate that our model has significantly lower computation runtime per image than the baselines, both at the mobile side, as well as the total computation time.

\Section{Related Works}

\subsection{Analytics-aware Image Compression}

One approach to computation offloading is to compress  the captured image at the mobile device and send the compressed bitstream for analysis at the edge server. But lossy compression of an image will invariably result in artifacts that degrades analytics performance. There have been multiple works which proposed methods to alleviate this loss in performance. 

A task-aware JPEG compression model was proposed in \cite{choi2020task}. The work proposes to use a convolutional network to predict the quantization table for DCT coefficients for JPEG compression. Images compressed using the predicted quantization table achieved better performance than when the standard table was used. In \cite{chamain2021end, le2021image}, the learned image compression model with hyperprior \cite{balle2018variational} was used to compress images then feed the compressed image to a detection or segmentation model. By end-to-end joint training of the compression model and the task model, some of the performance loss due compression is recovered. These prior works  require large convolutional networks to  operate on the mobile device, which still require substantial computation time and battery consumption. 

\subsection{Analytics-aware Feature Compression}

Another approach to computation offloading is to split the computation required for the analytics task between the mobile device and the edge server. The intermediate feature at the point of split is compressed at the mobile side and sent to the edge server. Intermediate feature compression using standard image/video codec was studied in \cite{chen2019toward, duan2020video} with HEVC, and in \cite{eshratifar2019bottlenet} with JPEG and additional dimensionality reduction. Using standard codecs for feature compression in general did not achieve good performance, as the codecs were designed for compression of images and not features.
Feature compression with learned image compression model was proposed in \cite{singh2020end}. The hyperprior compression model was used and the network was end-to-end trained. However, the split point considered by \cite{singh2020end} was at the very end of the original deep learning model, so that the mobile device still has to do majority of the computation task. 
In \cite{matsubara2022supervised}, knowledge distillation was used to teach a light-weight student network to compress the intermediate features of a large teacher detection network.

\begin{figure*}[!htb]
    \centering
    \includegraphics[width=\linewidth]{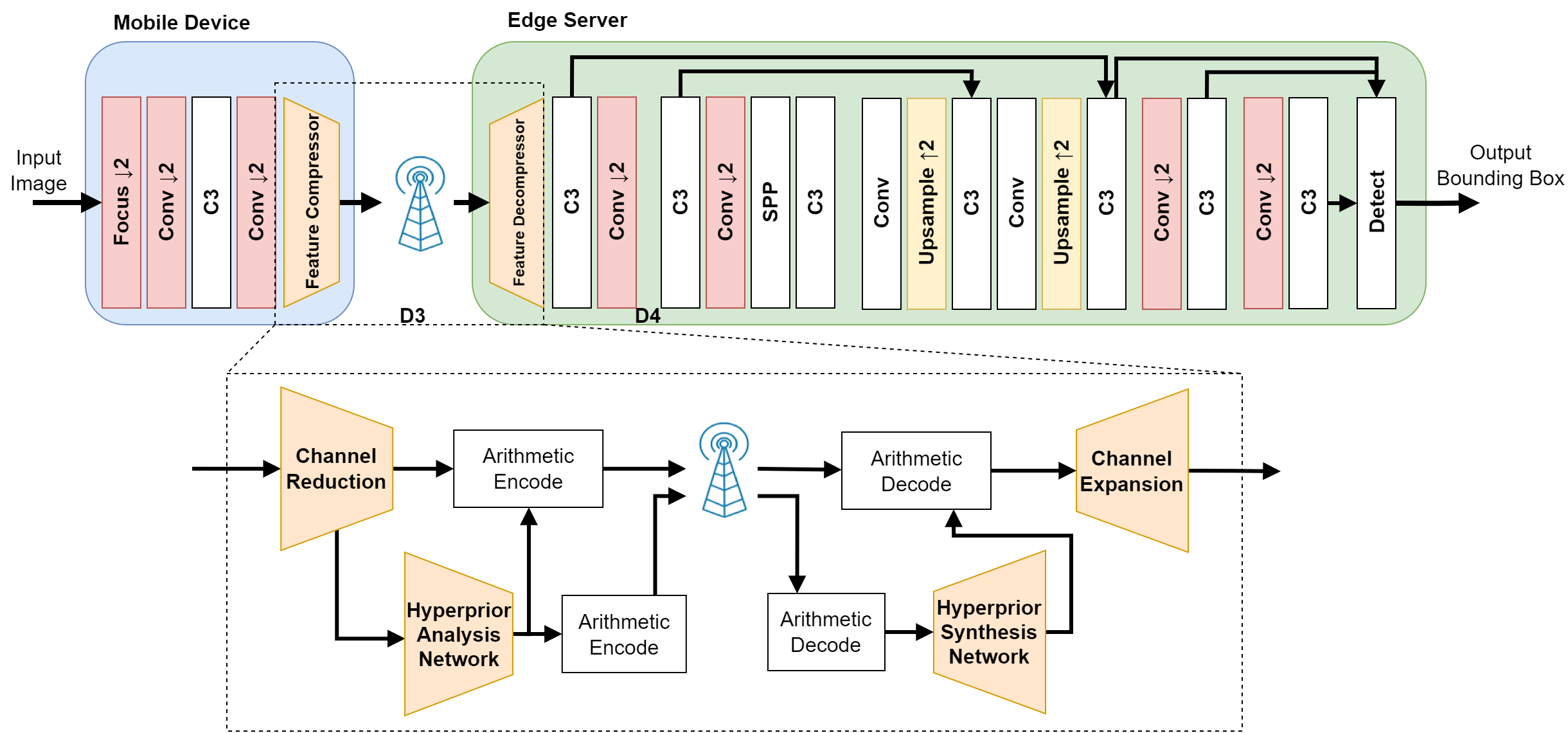}
    \caption{Overview of the proposed system. The YOLO network is split into two parts that separately runs on the mobile device and the edge server. A feature compression model is used to compress the intermediate features at D3. Another split point (D4) considered in our experiments is also labeled. When splitting at D4, the first skip connection of  the YOLO model has to be removed. The notation ↓2 and ↑2 refers to downsample by a factor of 2 and upsample by a factor of 2 respectively.}
    \label{fig:system}
\end{figure*}

\Section{Method}

\subsection{The Overall System Architecture}
\vspace{-2mm}
We propose a split computing system to offload computation for the YOLO detection model. We split the YOLO model into two parts: the first part runs on the mobile device and the second part runs on the edge server. The feature maps generated by the first part will be compressed by a feature compressor, also running on the mobile device. At the edge server,  the received bits are decoded by a feature decompressor and then sent to the second part of the detection model. See Fig.~\ref{fig:system} for the overall system architecture.

To choose the point of split we have to consider the trade-off between the mobile-side computation complexity and the compressed feature rate. In general, feature maps that are later in the detection model becomes more sparse and easier to compress. However, compressing later into the model requires more computation performed by the mobile device, defeating the purpose of split computing. Furthermore, in a pyramidal model architecture like the YOLO, shown in Fig.~\ref{fig:system}, there are skip connections which connect various layers in the ``Backbone'' to the ``Head" of the model. Compressing features after a skip connection would require removal of that skip connection. This can limit the detection performance, even though the bit rate is lower. 
%
%
We experimented with two points of split: 1) after the 3rd downsampling layer (designated as D3), and 2) after the 4th downsampling layer (D4), as illustrated in Fig.~\ref{fig:system}. Since the 4th downsampling layer is after a skip connection, we first removed the skip connection and finetuned the detection model before joint training for rate-constrained detection. We note that this has the effect of removing the high resolution information into the object detection branch, reducing the detectability of very small objects.  

In our feature compressor, we first employ a channel reduction layer to both reduce the spatial dimension and the number of feature channels and to decorrelate the remaining channels. 
We further adopt the hyperprior idea of \cite{balle2018variational}, to assist the entropy coding of the quantized feature maps after the channel reduction layer. The decoded hyperprior features are used  to estimate the probability distribution parameters of the feature maps. Both the bits for the quantized main features and those for the quantized hyperprior features are sent to the edge server.
The general architecture of feature compressor and feature decompressor is illustrated in Fig.~\ref{fig:system}.

\subsection{Channel Reduction and Expansion}
\vspace{-3mm}
The YOLO model uses a pyramidal architecture to extract features at different scales. At each downsampling layer, the spatial dimension is reduced and the channel dimension is increased. While having a large number of channels help improve object detection, it creates redundancy among the channels which can be seen in the inter-channel correlation matrix, as shown in Figure \ref{fig:cov}.

To reduce the inter-channel correlation, we introduce a channel reduction layer in the feature compressor, and correspondingly a channel  expansion layer in the feature decompressor. 
The channel reduction layer  reduces the number of channels  from $N$ to $N_r < N$. The channel expansion layer  takes the compressed features and expands it back to $N$ channels. The channel reduction is implemented by a $1 \times 1$ convolution layer, which  transforms $N$ features at each pixel to $N_r$ features. The channel expansion is implemented by a reversal $1 \times 1$ convolution layer, mapping $N_r$ features at each pixel to $N$ features. The channel reduction layer does not use nonlinear activation, while the channel expansion layer uses the SiLU activation as in the original YOLO model. This is because we want the reduced features to follow a Gaussian distribution to facilitate the entropy coding, while the SiLU activation would produce a single-sided feature values distribution.
If the split point is D3, we also employ a stride-2 convolution layer to reduce the spatial resolution, before  the $1 \times 1$ convolution layer for channel number reduction.

\subsection{Rate-constrained  Model Training}

In the learned compression framework proposed by \cite{balle2016end},  an input image ${x}$ is transformed into a latent feature ${y}$, which depend on ${x}$ and model parameter ${\theta}$. We express this dependency by writing ${y}$ as ${y({x};\theta)}$. The quantized latent feature ${\hat{y}}$ is decoded back to a reconstructed image ${\hat{x}(\hat{y};\theta)}$. The compression model is trained through minimizing a rate-distortion loss,
\begin{equation}\label{eq:RD-MSE}
\begin{aligned}
L & = L_{R} + \lambda \cdot L_{D} \\
L_{R} & = \displaystyle \E_{{x}\sim p_x}[-\log_{2}p({\hat{y}({x};\theta))}] \\
L_{D} & =  \displaystyle \E_{{x}\sim p_x}[d({x},{\hat{x}(\hat{y};\theta)})]
\end{aligned}
\end{equation}
where $\lambda$ is a hyper-parameter that controls the rate-distortion trade-off, ${\hat{y}}$ is the quantized latent, and $d({x},{\hat{x}})$ is the distortion between the original image ${x}$ and the decoded image ${\hat{x}}$. The distortion metric $d(\cdot,\cdot)$ for image compression is typically Mean-Squared-Error (MSE) or MS-SSIM.

In our framework, ${y}$ is the feature generated at the split point of the YOLO network, which is further reduced to ${z}$ by the channel reduction layer. We use ${\theta}$ to denote YOLO model parameters, and use  ${\phi}$ to indicate the channel reduction and expansion layer parameters. Thus, we write ${{z}(y(x;\theta);\phi)}$  to indicate these  dependencies.  We modify the rate-distortion loss in Eq.~(\ref{eq:RD-MSE}) to perform end-to-end training of the entire system including the YOLO detection model (parameterized by $\theta$) and the feature compressor and decompressor (parameterized by $\phi$) inserted at the split point, for detection-aware compression. One approach of training would be to use a distortion measure, such as MSE, between the original feature ${y}$ and the decompressed feature ${\hat{y}}$. However, minimizing such distortion  may not bring optimal detection performance under a constrained bit rate. Instead, we replace the distortion loss by a detection loss $L_{det}$ that directly measures the detection accuracy of the model's output:
\begin{equation}\label{eq:RD-Det}
\begin{aligned}
L & = L_{R} + \lambda \cdot L_{det} \\
L_{R} & = \displaystyle \E_{{x}\sim p_x}[-\log_{2}p({\hat{z}(y(x;\theta);\phi)})] \\
L_{det} & =  L_{obj} + L_{class} + L_{box}.
\end{aligned}
\end{equation}
$L_{det}$ is the loss used for training the uncompressed YOLO model, and it consists of the object detection loss $L_{obj}$, object class loss $L_{class}$, and bounding box loss $L_{box}$, which will depend on ${\hat{y}(\hat{z};\phi)}$ and $\theta$. A combination of rate and detection loss allows us to perform end-to-end training of the entire model including both the compression and detection components.

Instead of directly performing entropy coding on the quantized version of the reduced feature ${\hat{z}}$, we follow the hyperprior idea proposed in \cite{balle2018variational} to generate the hyperprior feature that help the entropy coding of  ${\hat{z}}$.
As shown in Fig~\ref{fig:system}, the hyperprior encoder generates quantized hyperprior feature  ${\hat{z}}_{h}$, from  ${\hat{z}}$. The hyperprior decoder predicts the mean and variance of each element in ${\hat{z}}$, in the Gaussian model used for entropy coding of  ${\hat{z}}$. The additional rate for ${\hat{z}}_{h}$ is included in the rate loss $L_{R}$.


Through our experiments, we find that it is necessary to perform a pre-training step of the channel reduction and expansion layers, along with YOLO model,  using only the detection loss. This step does not invoke quantization (through adding random noise) on the reduced feature $z$, nor the hyperprior encoder and decoder. 
Pre-training allows the model to reach an acceptable detection performance from the reduced feature channels before training with the compression objective, which involves adding noise to the reduced features  that can negatively affect detection performance.

\Section{Results}
\vspace{-2mm}

\subsection{Experimental Settings}
\vspace{-2mm}

We adopted the Ultralytics YOLOv5 implementation to perform our experiments \cite{YOLOv5}. We chose the smaller model size YOLOv5s with 7.2M parameters for faster training and inference. When performing pretraining to the models without rate constraint, we initialized the model with weights from \cite{YOLOv5}, which were trained with the images in the entire COCO dataset (training set only). All the original images were resized to $640 \times 640$. All models were trained using SGD with a learning rate of $1\times10^{-4}$ and a momentum of 0.937. We used a batch size of 96 when training the traffic-specific models and a batch size of 120 when training the full COCO dataset models. We evaluated the achievable performance by varying the split position (D3 vs. D4), the number of channels after channel reduction ($N_r$), and $\lambda$ in the loss function, for a chosen split position and $N_r$. We show the Pareto front of these various settings that achieves the highest detection performance under the same bit rate.

We compare the performance of the proposed approach with the following benchmarks: 1) Using the popular BPG image compression algorithm \cite{BPG} to compress the input image, and then applying the original YOLO model for object detection. The BPG implements the HEVC intra coding algorithm and provides leading compression performance among standard public codecs; 2) Using the pretrained learned compression model of \cite{balle2018variational} to compress the input image, followed by  the original YOLO model for object detection; 3) Training the learned compression model of \cite{balle2018variational} and the YOLO model jointly, using the same detection-aware loss function in Eq.~(\ref{eq:RD-Det}). 

\subsection{Detection vs. Rate Performance using the Full COCO Dataset }
\vspace{-2mm}

\begin{figure}
    \centering
    \includegraphics[width=\linewidth]{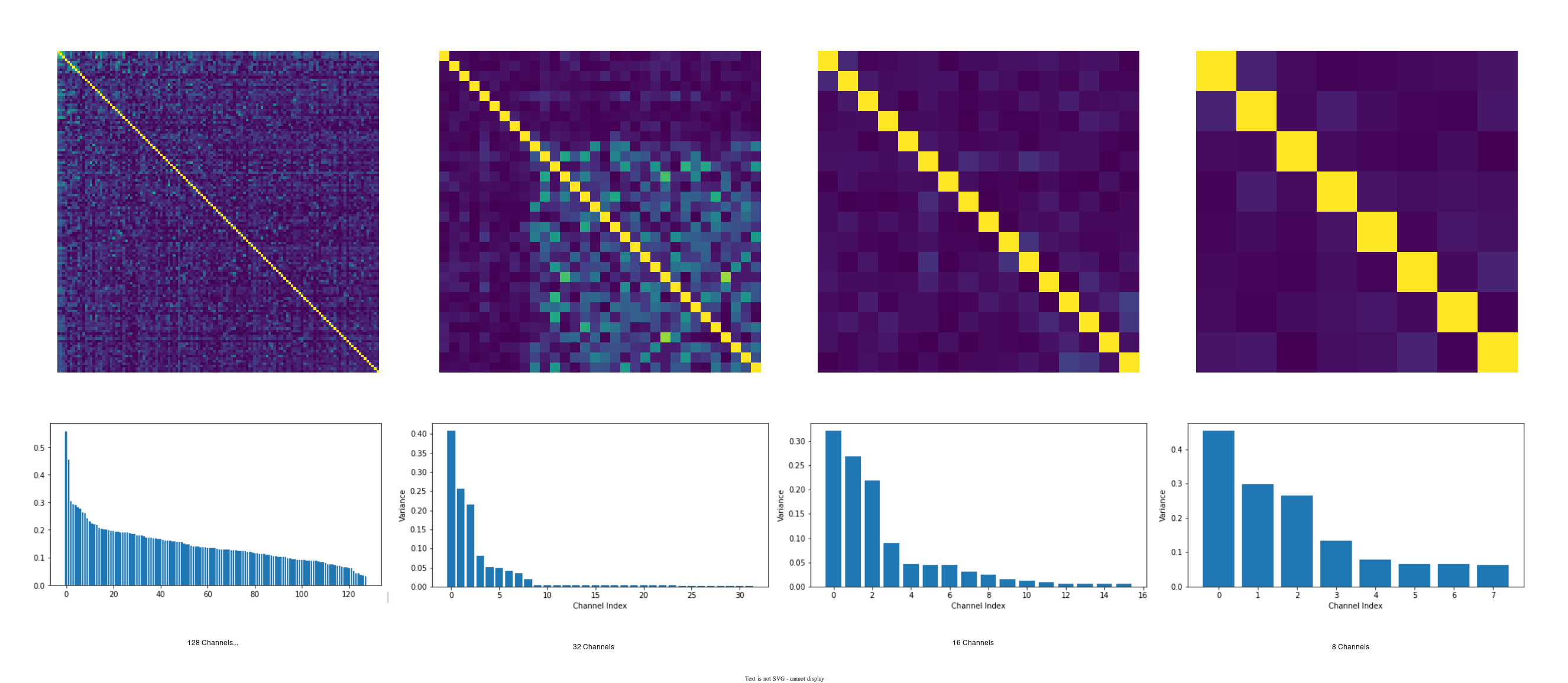}
    \caption{Inter-channel correlation matrix (top row) and the variance of each channel (bottom row) for the intermediate feature map at the D3 split point.}
    \label{fig:cov}
\end{figure}

First, we show that the channel reduction module was able to reduce the correlation between channels. As shown in Figure \ref{fig:cov}, the off-diagonal elements of the correlation matrix becomes more sparse as the number of channels after reduction decrease, which means there is less correlation among channels. Secondly, the variance plots show that only a few channels have a large variance magnitude. This suggests that it is possible reduce the number of feature channels to a small value without losing much information in the feature.
    
\begin{figure}[!ht]
     \centering
    \includegraphics[width=\linewidth]{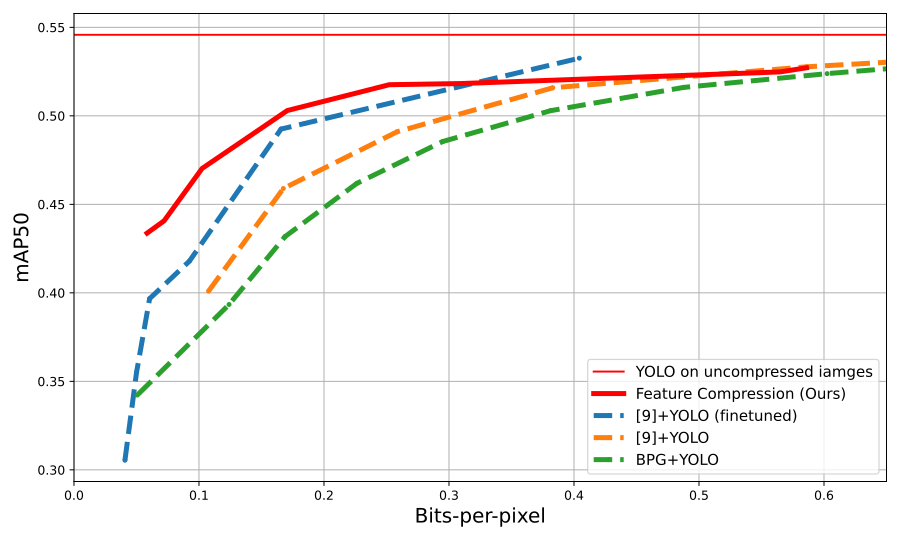}
    (a) Full COCO Dataset (80 object classes)
    \includegraphics[width=\linewidth]{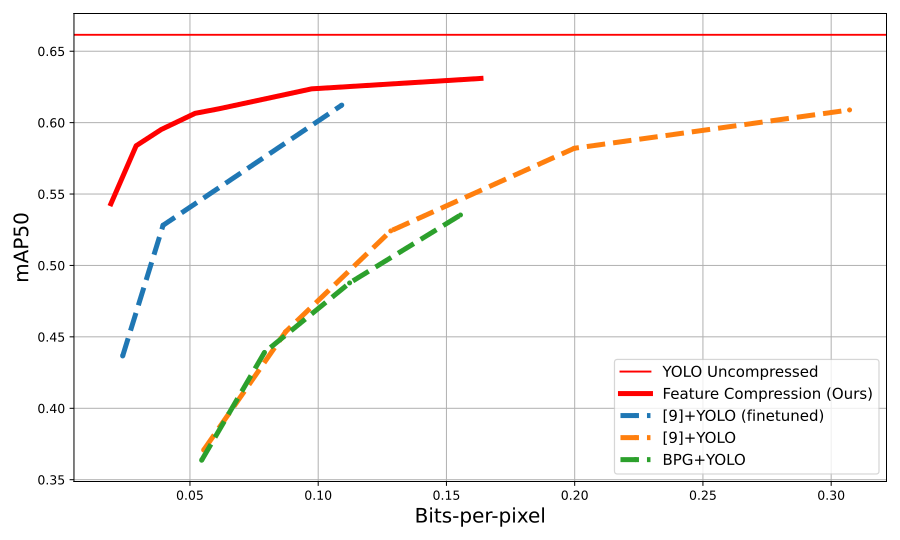}
    (b) COCO-Traffic Datset (9 object classes)
    \caption{Detection performance under various bitrates for the full COCO dataset and the COCO-Traffic dataset.}
    \label{fig:performance}
\end{figure}

The rate vs. detection accuracy curves of our feature compression model and the benchmark methods are shown in Figure \ref{fig:performance} (a). The detection accuracy is measured by the mean Average Precision under the Intersection over Union threshold of 50 (mAP50).
%
%
%
%
Our method performed better than the baselines of image compression  followed by YOLO detection over the entire rate range considered, although the difference is small at the higher rates. Compared to jointly training the image compression model \cite{balle2018variational} and the YOLO model for rate-constrained object detection, our method is still better at the lower rate regime. We expect that with  more exhaustive search of the hyper parameters in our feature compression layer, our method can be on-par with this benchmark over the higher rate range.

\subsection{Performance using the COCO-Traffic Dataset}
\vspace{-2mm}
The original YOLO model was trained with the COCO dataset \cite{COCO}, which aims to detect 80 classes of objects from various environments. However, in many practical use cases, it is not necessary to consider such a wide variety of classes, but rather to focus on detecting a small set of objects. We hypothesize that in such a targeted use case, we can significantly reduce the bit rate without sacrificing the detection accuracy. Specifically, we consider  a situation where only object classes that are relevant for traffic surveillance or navigation applications need to be detected.
From the COCO dataset we picked 9 relevant classes, including Person, Car, Bus, Truck, Motorcycle, Traffic Light, Fire Hydrant, Stop Sign, Parking Meter.
All images with occurrence of at least one of these 9 classes are drawn into a dataset called the COCO-Traffic dataset. Since the person class appears in many images with non-traffic related scenarios, we additionally require images containing the person class to be coupled with at least one other traffic-related classes for those images to be drawn. Using the COCO-Traffic dataset, we perform the same experiments as for the full COCO dataset and the results are shown in Figure \ref{fig:performance}(b).


For this experiment, we only evaluated models at the D4 split point and focused on the  low bitrate region. By focusing on a more specific set of objects, the features can be compressed to very low bitrates while still maintaining high mAP. The results are significant for settings with low communication bandwidth and require compression into extremely low bitrates. It is possible to deploy a split detector in these settings to have high detection performance while using low bandwidth transmission.


\renewcommand{\arraystretch}{0.9}
\begin{table*}[htbp]
\resizebox{\textwidth}{!}{\begin{tabular}{l|lcc|lcc|c}
\hline
                                                                               &                                      & \cite{balle2018variational} + YOLO           & BPG + YOLO              &                                        & C4\_D3                 & C4\_D4                 & YOLO on Mobile        \\ \hline
\multirow{3}{*}{\begin{tabular}[c]{@{}l@{}}Mobile Device\\ (CPU)\end{tabular}} & \multirow{3}{*}{Image Compression}   & \multirow{3}{*}{415.46} & \multirow{3}{*}{183.87} & YOLO Pre-split                         & 55.18                  & 67.37                  & \multirow{3}{*}{2791} \\
                                                                               &                                      &                         &                         & \multirow{2}{*}{Feature Compression}   & \multirow{2}{*}{8.32} & \multirow{2}{*}{8.88}  &                       \\
                                                                               &                                      &                         &                         &                                        &                        &                        &                       \\ \hline
\multirow{3}{*}{\begin{tabular}[c]{@{}l@{}}Edge Server\\ (GPU)\end{tabular}}   & \multirow{2}{*}{Image Decompression} & \multirow{2}{*}{1.82}   & \multirow{2}{*}{84.39}  & \multirow{2}{*}{Feature Decompression} & \multirow{2}{*}{0.27} & \multirow{2}{*}{0.25} & \multirow{3}{*}{0}    \\
                                                                               &                                      &                         &                         &                                        &                        &                        &                       \\
                                                                               & YOLO                                 & 7.39                   & 7.39                   & YOLO Post-split                        & 6.60                  & 5.30                  &                       \\ \hline \hline
Total time on Mobile                                                           &                                      & 415.46                  & 183.87                  &                                        & 63.50                  & 76.25                  & 2791                  \\
Total time on Server                                                           &                                      & 9.207                   & 91.78                   &                                        & 6.87                   & 5.55                   & 0                     \\ \hline
\end{tabular}}
\caption{Breakdown of runtime (milliseconds) per image ($640\times 640$ pixels) for proposed model and baselines. C4\_D3 and C4\_D4 refer to models with channel reduction to $N_{r}=4$ and split point at D3 and D4, respectively. The results here are similar regardless of bitrate and the channels $N_r$.}
\label{table:runtime}
\end{table*}

\subsection{Mobile and edge computing time}
\vspace{-2mm}
We performed an analysis of the runtime for each model under a mobile-edge split computing setting. 
For the mobile device we used a 2.9GHz CPU processor to perform the computations from input image to encoded bit-stream. For the edge server, we used the powerful RTX8000 GPU to decode the received bitstream and run the detection network. 

As shown in Table~\ref{table:runtime}. Our method has a clear advantage against the baselines in both the mobile computation time as well as the total runtime. For example, with splitting at D3, compared to using the learned image compression model \cite{balle2018variational}, the mobile computing time is reduced by 6.58x. The server computing time is also reduced, as  part of the YOLO model is already executed by the mobile, resulting a total runtime saving of 6.06x. Compared to using the BPG image coder, the mobile runtime is reduced by 2.91x, while the total computation time is reduced by 3.93. It turned out that GPU is not efficient for  running the BPG decoder, which requires sequential operation.   The proposed method is the only solution that can enable object detection at a speed that is faster than 10 frames per second (requiring total runtime $\leq$ 100 ms), required for most practical applications. 


\Section{Conclusions}
\vspace{-2mm}
This paper proposed an approach for offloading deep-learning based object detection by split computing between the mobile device and the edge server. We propose a light-weight trainable feature compression and decompression architecture, that includes feature channel reduction/expansion and hyperprior-based entropy coding/decoding.   With end-to-end training of the feature compressor and object detector using rate-detection loss, our approach can achieve higher detection accuracy at low to medium  rate range, than baseline methods that perform image compression at the mobile device and object detection on the server.  Furthermore, our approach has significantly lower runtime at the mobile device than the baseline methods.
\nocite{ex1,ex2}
\bibliography{latex8}

\end{document}